\documentstyle[12pt,epsf]{article}

\newcommand{\be}{\small\begin{equation}}
\newcommand{\ee}{\end{equation}\normalsize\vspace*{-0.1ex}}
 \newcommand{\bea}{\small\begin{eqnarray}} 
\newcommand{\eea}{\end{eqnarray}\normalsize\vspace*{-0.1ex}}
 \newcommand{\bdm}{\small\begin{displaymath}} \newcommand{\edm}
{\end{displaymath}\normalsize\vspace*{-0.1ex}} \newcommand{\beas}
{\small\begin{eqnarray*}} 
\newcommand{\eeas}{\end{eqnarray*}\normalsize\vspace*{-0.1ex}}

\begin{document}


\thispagestyle{empty}
\renewcommand{\thefootnote}{\fnsymbol{footnote}}

\setcounter{page}{0}
\begin{flushright} UM-TH/97-18\\
hep-ph/9710487
\end{flushright}

\begin{center}
\vspace*{1.6cm}
{\Large\bf
On Non Perturbative Corrections to the Potential for Heavy Quarks}

\vspace{1.5cm}
 R. Akhoury and  V.I.Zakharov
\vspace{0.6cm}

{\it Randall Laboratory of Physics\\
University of Michigan\\ Ann Arbor, Michigan 48109-1120}\\[0.6cm] 

{\bf Abstract}\\[0.6cm]
\end{center}

We discuss non perturbative corrections to the Coulomb-like potential
 of heavy quarks at short distances.
We consider both the standard framework provided by infrared
 renormalons and the assumption that confinement does not allow weak fields to
 penetrate the vacuum. In the former case the leading correction at short
 distances turns out to be quadratic in $r$
 for static quarks. In the latter case
 we find a potential which is proportional to $r$ as $r \rightarrow 0$.
We point out that similar effects arise due to a new kind of non perturbative
 correction proportional to $1/Q^2$, which is unaccounted for by the operator
 product expansion and which was recently discussed within a different
framework.
 Phenomenological implications of the linear correction to the 
potential are briefly
reviewed.

\vspace{0.8cm}
\noindent PACS numbers: 12.38.Cy, 12.39.Hg, 13.20.He

\newpage
\renewcommand{\thefootnote}{\arabic{footnote}} \setcounter{footnote}{0}


{\bf 1.} Analytic studies of non-perturbative effects in QCD are
 limited by the fact that there does not exist a systematic procedure
to account for confinement. We may attempt, however, to sort out and to 
parametrize
 its effects by considering the case when the perturbative part is dominant
 and the infrared sensitive contributions appear only as small power corrections
 in a large mass scale $Q$.
Well known examples of this kind are provided by the applications of the operator
 product expansion such as the QCD sum rules \cite{svz}. More recently,
 renormalons \cite{thooft} have emerged as a more universal tool for
 enumerating the power-like corrections even in situations where the
operator product expansion does not apply ( for furthur references and
 review, see e.g., \cite{review}).

In this note we will apply these techniques to the short-distance
interaction between heavy quarks, $Q, \bar{Q}$. As is well known,
 this interaction is dominated by a Coulomb-like potential:
\be
\lim_{r\rightarrow 0} V(r)~=-{C_F\alpha_s(r)\over r} \label{potential} \ee
where the coefficient $C_F=4/3$ refers to the color-singlet channel.
We will argue below that the character of the leading power corrections
 depend on the properties of confinement.
In particular, the standard renormalon technique
implies the effect of confinement to become important only at distances
 $R_{cr} \sim \Lambda_{QCD}^{-1}$ and the leading power correction to
(\ref{potential}) turns to be:
 \begin{equation}
\lim_{r\rightarrow 0}\delta V(r)_{renormalon}~=~c_2 r^2. \label{renormpot} 
\end{equation}
Similar results for the static potential have already been obtained in
the literature by various techniques \cite{balitsky,simonov,aglietti}. 

Although infrared renormalons probably correctly indicate the presence
 of some of the power corrections, it is an open questions whether they
 exhaust all the possibilities. 
In case of the static potential,
infrared renormalons do not give any indication to a formation
of a string at large distances and could miss, therefore, 
the leading correction at short distances as well. In fact, we will 
consider the power
 corrections to the static potential (\ref{potential}) in an alternative
 model which implements the assumption that the effects of confinement
 become important once the color fields become weak. In such a case
 (see below ) we find in contrast to Eq. (\ref{renormpot}), a leading
 correction which is linear in r: 
\begin{equation}
\lim_{r\rightarrow 0}\delta V(r)_{non-standard}~=~c_1 r \label{linear} 
\end{equation}
A crucial element to ensure such corrections appears to be the existence 
of small size non perturbative fluctuations in QCD.
 
Power corrections which go beyond the OPE and infrared renormalons have
 recently been highlighted both within a
general dispersion approach \cite{grunberg,shirkov} and
the ultraviolet-renormalon technique \cite{vz}. Most remarkably,
indications from lattice simulations for the existence of such corrections
 have very recently been obtained in \cite{marchesini}. 
More specifically,
 one argues that the effective coupling may contain terms proportional to
$1/Q^2$:
\begin{equation}
\alpha_{eff} (Q^2)~=~\alpha_{eff}^{pert.}(Q^2)~+~c_{Q^2}{\Lambda_{QCD}^2 \over Q^2}
 \label{alpha} \label{novel}\end{equation}
while infrared renormalons only indicate the presence of corrections
 proportional to $1/Q^4$, corresponding to the gluon condensate.

 It is amusing to observe that the $1/Q^2$ piece in
 (\ref{alpha}) also would give
 rise to a linear correction to the potential. Thus it is interesting to
 speculate that the mechanism of the repulsion of weak fields by the vacuum
could be a possible mechanism responsible for the correction (\ref{alpha})
 if its existence \cite{marchesini} is confirmed. 
We would like to emphasize that although the static potential
 is not directly observable, it can be measured on the lattice (for a recent
 study including further references see \cite{bali}). In particular,
 although the existence of a linear potential at {\it large } distances
is well established and there is no indication 
to its change to the $r^2$ behaviour as predicted by (\ref{renormpot}), 
much more precise data is needed
to either confirm
 or reject the presence of a linear term at {\it short} distances. On the
 other hand the linear correction Eq. (\ref{linear}) would also affect 
the by-now-standard
 theory of the bound states of heavy quarks (for a review and further
 references see, e.g., \cite{yndurain})
and we will address this issue as well.

{\bf 2.} To evaluate the renormalon contribution to $V(r)$ consider first the
 one gluon exchange potential
\be
V(r)~=~-C_F\int d^3{\bf k}(4\pi \alpha_s({\bf k}^2))
{\exp(i{\bf k\cdot r})\over {\bf k}^2} \label{fourier} \ee
where the leading logarithmic corrections have been incorporated into
 the running coupling constant, $\alpha_s({\bf k}^2)$, and the fact that
for static quarks it depends on ${\bf k}^2$ is crucial for further argument.
Beyond the leading logs, 
Eq. (\ref{fourier}) holds in an Abelian case and in the limit
of large $N_f$ in QCD \cite{aglietti}. 
 To find the renormalon contribution we write \be
\alpha_s({\bf k}^2)~=~\int d\sigma \left({{k}\over
{\Lambda_{QCD}}}\right)^{-2\sigma b_0}
 \label{renorm} \ee
whee $b_0$ is the first coefficient in the expansion of the $\beta$-function
 (for simplicity we confine ourselves to a one-loop $\beta$-function), and
 $\Lambda_{QCD}$ is the position of the Landau pole.

Next, we substitute (\ref{renorm}) into (\ref{fourier}) and perform the
integration over directions of ${\bf k}$ to get \be
V(r)~=~-{16\pi\over 3}\int_0^{\infty}d\sigma\int_0^{\infty}dk {\sin
 (kr)\over kr}
 \left({k\over \Lambda_{QCD}}\right)^{-2\sigma b_0} .\ee
Moreover, and since we are interested in $V(r)$ at small $r$, we expand in
$kr$: \be
V(r)~=~-{16\pi\over 3}\int_0^{\infty}
\Lambda_{QCD}^{2\sigma b_0} d\sigma\int dk (k^{-2\sigma b_0}+
k^{2-2\sigma b_0}r^2+...),
\ee
The renormalon poles at $\sigma =1/2b_0, 3/2b_0 ...$ are clearly seen in the
 above and this suggests a modification of the Coulomb potential at small
$r$ of the form:
\be
\delta V(r)~=~c_0\Lambda_{QCD}~+~c_2\Lambda_{QCD}^3r^2+...\label{modified}
 \ee
where $c_{0,2}$ are constants and it is noteworthy that they do not contain
 any small coupling.  This has already been noted in \cite{aglietti}. Since
 we expand in $kr$ and $k\sim\Lambda_{QCD}$ at the renormalon poles,
 Eq. (\ref{modified}) is valid at distances small compared to the
confinement radius: \be
r~\ll~\Lambda_{QCD}^{-1}. \label{less}
\ee
This result has been obtained via a Borel transformation to conform to
 the current usages in renormalon analysis. It may also be seen by
 noting that the pole at $k=\Lambda_{QCD}$ in the standard, leading
 log expression for $\alpha_s$,
\be
\alpha_s({\bf k}^2)={1 \over{{b_0}\log k^2/\Lambda_{QCD}^2}} \ee
implies that the Fourier integral above is undefined. According
 to the general theory of singular integrals, we may write
\be
\alpha_s({\bf k}^2)={\rm P.P.}{1\over{{b_0}\log k^2/\Lambda_{QCD}^2}}+
 {\rm i}c\Lambda_{QCD}^2\delta(k^2-\Lambda_{QCD}^2), \label{landau} \ee
where $c$ is an arbitrary constant. When integrated the last term,
 ${\rm i}c\Lambda_{QCD}^2\delta(k^2-\Lambda_{QCD}^2)$, is easily
 seen to give a correction to the potential, \be
\delta V(r)~=~c{\sin (r\Lambda_{QCD})  \over r} \label{sin}\ee
which, upon expansion, reproduces the type of corrections given in Eq.
(\ref{modified}).

Eq. (\ref{modified}) remains qualitatively unchanged upon inclusion
of higher loop corrections as well. 
The results of explicit calculations (which are known now
up to three loops \cite{peter}) are usually represented as an expansion,
in $\alpha_{\overline{MS}}$:
\be
V({\bf k}^2)~=~-C_F{4\pi\sum a_n\alpha^n_{\overline{MS}}({\bf k}^2)\over
{\bf k}^2}.\label{expansion}\ee
It is straightforward to see that renormalon contribution
associated with higher powers of $\alpha ({\bf k}^2)$ give the same type
of power corrections as above.
Indeed, in the approximation of the one-loop $\beta$-function,
\be
\alpha^2({\bf k}^2)~=~{1\over 2b_0}\Lambda_{QCD}{d\over \Lambda_{QCD}}
\alpha({\bf k}^2)
.\ee
Applying the differentiation with respect to $\Lambda_{QCD}$ at the last step,
i.e. to Eqs. (\ref{modified},\ref{sin}) we immediately see that the Fourier transform
of the $\alpha^2 ({\bf k}^2) $ term brings the same
kind of power corrections as above. This is true for higher powers of
$\alpha ({\bf k}^2)$ as well so that Eq. (\ref{modified}) is reproduced
by renormalons associated with any order in $\alpha ({\bf k}^2)$
in Eq. (\ref{expansion}). Finally, inclusion
of higher loops in the $\beta$-function itself is known \cite{mueller}
to modify the renormalon poles to renormalon cuts, the position of the
singularity remaining unchanged. This would bring powers of logs in
the correction to the potential but would not change the power
of $r$, $r^2$.

It is worth pointing out that the renormalon contribution
 could be tried also at large distances. The resulting potential does not
 reproduce the linear rise at large $r$ and is readily seen to be devoid of
 physical meaning. The reason is that renormalons are a pure perturbative
 construct and do not produce any hint that at large $r$ a string is formed. 
This failure demonstrates a limitation of the renormalon technique -- the existence of
 the Landau pole in the infrared region does not automatically imply confinement.
 
To go beyond the perturbative expansion (\ref{expansion}) 
for $V(r)$ in the 
non- Abelian case one has to consider the Wilson loop average $<W(C)>$ for
the stretched rectangle $C=r\times T$ with small $r$ and large $T$:
\be
V(r)~=~-\lim_{T\to\infty}{1\over iT}ln
<Tr~Pexp\left(ig\oint_Cdx_{\mu}A^{\mu}_aT^a\right)>
.\ee
While there is no rigorous way to evaluate $<W(C)>$ analytically,
model calculations \cite{balitsky,simonov} do reproduce 
the same $r^2$ correction as the leading one at short
distances.
In particular
a model of Ref. \cite{balitsky} gives the following result:
\be
V(r)~=~-C_F{\alpha(r)\over r}+r^2\Phi (r)+O(r^4),
\ee
where 
\be
\Phi (r)~=~{4\pi\over 72} <\alpha_s(G_{\mu\nu}^a)^2>
\left(\rho^{-1}+{3\over 2}r^{-1}\alpha_s(r)\right)^{-1}\label{rho}
\ee
and $\rho$ is the characteristic size of non-perturbative fluctuations
dominating the gluon condensate, $ <\alpha_s(G_{\mu\nu}^a)^2>$.
Thus, we have the same $r^2$ correction in so far as $$<\alpha_s(G_{\mu\nu}^a)^2>
\sim\Lambda^4_{QCD}, \rho\sim\Lambda^{-1}_{QCD}$$.

{\bf 3}. The examples given above indicate that it is the size of 
non-perturbative
fluctuations which is crucial to determine  
the character of leading power corrections to the potential at short
distances.
What unifies renormalons and the model underlying Eq. (\ref{rho}),
is that in both cases the characteristic size of non-perturbative
fluctuations is of order $\Lambda^{-1}_{QCD}$.
To visualize the connection between the short and large distances
in the most transparent way let us
represent, in an Abelian case,
 the potential energy of the $Q\bar{Q}$ pair as an integral over space from 
the
quark electric fields: \be
V(r)~=~{1\over 4\pi}
\int d^3r'{\bf E}_1({\bf r}')\cdot{\bf E}_2({\bf r+r'}). \ee
In particular, the Coulomb potential can be obtained of course by 
integrating
over the fields ${\bf E}_{1,2}$ of two point charges. On the other hand, if
the electric fields are modified at large distances, then there arises a correction
to the Coulomb energy at small distances as well.

Consider as an example two charges of opposite signs in a cavity of size $R$.
 Then the electric field of the charges, which is that of a dipole at large
 distances in empty space, changes at $r'>R$. The corresponding change in
$V(r)$ is of order
\be
\delta V(r)~\sim~\alpha r^2\int_R^{\infty}
{d^3r'\over(r')^6}~\sim~{\alpha r^2\over R^3}\label{cavity} \ee
which is in agreement with the correction to the static potential discussed
above.
The renormalon-induced correction to the Coulombic potential
 (\ref{modified}) takes
into account that at $r'\ge \Lambda_{QCD}^{-1}$ the fields are distorted by 
the confinement effects.
The corresponding correction to the potential is of order \be
\delta V(r)~\sim~r^2\int_{\Lambda_{QCD}^{-1}}^{\infty}
 {d^3r'\over (r')^6}~\sim~r^2\Lambda_{QCD}^3 \ee
where $\alpha_s(r')\sim 1$ in this estimate. 

From this point of view, it is not at all obvious that the major
effect due to confinement is the modification of the Coulomb field of each
of the quarks at $r'\sim \Lambda_{QCD}^{-1}$. Consider
a simplified model according to which the electrostatic field of quarks
is a correct zeroth-order aproximation only as far as it exceeds some critical
value of order $\Lambda_{QCD}^2$: \be
({\bf E}^a)^2~\sim~\Lambda_{QCD}^4 \label{weak} \ee
while weaker fields do not penetrate the vacumm because of its specific,
confining
properties. From this condition we get an estimate of distances $R_{cr}$ where
 the electrostatic field of quarks is strongly modified: \be
{\alpha_s r^2\over R_{\rm cr}^6}~\le~\Lambda_{QCD}^4 \label{ratio}\ee
where for simplicity we have neglected the effect of the running of
$\alpha_s (r')$.

The change in the potential is then of order: \be
\lim_{r\to 0}\delta V~\sim~{\alpha_sr^2\over R_{\rm cr}^3}~\sim~\alpha^{1/2}r
\Lambda_{QCD}^2 \label{nnew},
\ee
i.e., we get a linear in $r$ leading correction to the potential at short
distances.
 Note that a linear potential in the context of models of a stochastic
vacuum has
 been claimed a long time ago \cite{dosch}, however, that derivation refers
to large distances
 while the estimate (\ref{nnew}) applies at short distances.
It is worth emphasizing, that the estimate (\ref{nnew}) is entirely
dependent on
the assumption (\ref{weak}) as applied to the dipole field of the
$Q\bar{Q}$ pair.
 Similarly, the renormalon effect can be visualized as arising from a similar
condition but applied to a Coulomb-like field of a quark. Since the field of
 a dipole is weaker than that of a charge, the critical value of ${\bf E}^2$
is reached at shorter distances and the feedback from these
distances is stronger.

It is worth emphasizing that to realize condition (\ref{weak})
with $r\rightarrow 0$ one needs small-size nonperturbative
fluctuations in the QCD vacuum. Indeed $R_{crit}\rightarrow 0$ as $r\rightarrow 0$.
This conclusion also fits well with the discussion above.
That is why a confirmation of the existence of the linear
correction (\ref{nnew}) would provide very important insight into the
mechanism of confinement. If one tries to speculate about what kind of fluctuations 
these could be, it is natural to turn to the dual superconductor
picture of confinement \cite{mandelstam} (for a recent review see
\cite{baker}). Magnetic monopoles are a crucial field configuration in this
case. The magnetic monopoles of QCD were introduced \cite{thooft1}
in the Abelian projection of QCD where they appear as singular objects.
Although this could be an artifact of the gauge fixing \cite{thooft1}
 convincing evidence for existence of monopoles as physical objects was
obtained in just this gauge (see \cite{polikarpov} and references therein).
If the physical size of monopoles is indeed vanishing,  the linear 
potential at large distances, or the area law for the Wilson loop,
could well continue to $r\rightarrow 0$ where it becomes a correction
to the Coulomb-like potential. Existing data on the lattice \cite{bali}
do not indicate any change in the linear in $r$ piece of the potential at small
distances but no special mesurements targeting this behaviour have been
performed so far.

As is mentioned above, very recent measurements do indicate 
\cite{marchesini} the presence
of a $1/Q^2$ correction in the  effective coupling (see Eq(\ref{novel})) with
a positive constant $c_{Q^2}$:
\be
(c_{Q^2})_{lat}~>~0\label{sign}
.\ee
We observe that this sign is in fact opposite to what one would expect
from a model in which the effective 
coupling at large distances is frozen\cite{shirkov}:
\be
\alpha_{fr}~\approx~{1\over b_0}\left ({1\over lnQ^2/\Lambda^2_{QCD} }
+{\Lambda^2_{QCD}\over \Lambda_{QCD}^2-Q^2}\right).\ee
On the other hand, if one assumes that linear in $r$ potential 
established at large distances continues to short distances as well
then the $1/Q^2$ correction to the coupling would have the same sign 
as indicated by the data (\ref{sign}). It is worth mentioning once more
that although most recent discussions \cite{grunberg,shirkov}
emphasize the analyticity aspect of the high-$Q^2$ behaviour 
of the coupling constant
the analiticity itself does not fix the form of the correction 
(see, e.g., \cite{dmw}). On the other hand, 
as is discussed above, the
existence of small-size fluctuations seems to
be a prerequisite for the $1/Q^2$ corrections.

{\bf 4.} So far we discussed the static quark potential which can be 
measured on a lattice.
Quark potential is also relevant to the physics of bound states of heavy quarks
(for a review  and references see \cite{yndurain}).
As the zeroth order approximation one considers usually Coulomb-like states
in the potential (1). Then
if one treats the potential $K^2r$ as a perturbation the corrections
to the energy levels are:
\be
\delta E_{nl}~=~{1\over 2}\big[3n^2-l(l+1)\big]aK^2, ~~~a~=~{2\over mC_F\alpha}
\ee
where $a$ is the corresponding Bohr radius.
Nominally this correction is enhanced 
by powers of $m/\Lambda_{QCD}$ as compared to correction due to
the possible $r^2$ piece in the potential in (see Eq. (\ref{modified}))
or the Voloshin-Leutwyler correction \cite{voloshin}.
In particular, the latter correction 
to energy levels is inversely proportional to $m^3$:
\be
(\delta E_1)_{VL}~=~{1.67\pi <\alpha_s (G^a_{\mu\nu})^2>\over m^3C^4_F\alpha^4_s}
\ee
where we have quoted the result for $n=1$ which might be the most
relevant to phenomenological applications.
There are, however, important caveats to this statement
on the dominating role of the corrections due to the linear term, if it exists.
First, standard calculations \cite{pantaleone} already contain a linear term
in the potential (extrapolated from large distances),
so that $K^2$ shoud be understood rather as a deviation from
this form of the potential at short distances. Furthermore, the
Voloshin-Leutwyler correction contains large factors, 
like $\alpha^{-4}$ which are important numerically. Since we do 
not know yet of an estimate of the constant $K^2$,
and also the value of the gluon condensate is not well known, 
it is difficult to compare
the relative size of various contributions. 
The best strategy at the moment for the detection 
of the short-distance $K^2r$ piece is, 
as already stated, through lattice
measurements.

{\bf 5}. On the theoretical side, the most important reservation about
applying and comparing various corrections simultaneously is that the
potential picture has a limited scope of validity. To clarify the applicability of
the renormalon-induced and of the linear correction (see Eqs. (\ref{modified}) and (\ref{novel}),
respectively) to quarkonium physics, let us first review the well known results from QED and QCD,

The particular QED effect which might imitate the interaction of quarks at
short distances with large scale fluctuations is the shift of atomic levels in a cavity.
In the static potential picture, the modification of the potential due to the electrostatic
interaction of a dipole with the cavity, $\delta V(r)\sim r^2/R^3$, is found first
(see Eq.(\ref{cavity}). The energy shifts could be
 obtained then by averaging this potential over the unperturbed wave functions.
This procedure would be equivalent to using the $r^2$ piece in the
potential (\ref{modified}) as a perturbation on the Coulomb-like
states.

A consistent quantum-mechanical treatment of the problem \cite{casimir}, on
 the other hand, starts with the dipole interaction: \be
H_{int}~=~ - e{\bf d\cdot E} \label{dipole}. \ee
where ${\bf E}$ is the electric field associated with zero-point fluctuations
of the electromagnetic field in the cavity. The spectrum of the zero-point
fluctuations depends on the size $R$ through the boundary
condition, and one is interested in fact only in these $R$-dependent terms.
Next, the shift in energy levels $E_n$ is obtained to second order in
 $H_{int}$: \be
\delta E_n~\sim~V_{nk}(E_n-E_k+\omega_{\rm char})^{-1}V_{kn}. \label{so} \ee
We consider bound states and have set
$V_{nk}=\langle n|\delta L_{int}|k \rangle $. Since we are interested only
in estimates,
 we simply retain the contribution of a characteristic photon frequency,
 $\omega_{\rm char}\sim 1/R$. 
The result for $\delta E_n$ depends crucially on the relative magnitude
 of $E_n \sim m\alpha^2 \sim \alpha/a$ and $\omega_{\rm char}$, i.e., on
 the relative magnitude of
$R$ and $a/\alpha$, where $a$ is the Bohr radius. Note that a new scale,
that is $a/\alpha$, emerges at this stage. Namely, if $\omega_{\rm char}
\gg E_n$ then
\be
\delta E_0~\sim~\alpha a^2 ({\bf E}^2)R~\sim~\alpha{a^2\over R^3}, ~~~~R\ll
a/\alpha
\ee
since $e^2{\bf d}^2 \sim \alpha a^2$.
 The shift corresponds of course
 to evaluating the matrix element of the electrostatic potential,
 $\delta V(r)\sim \alpha r^2/R^3$ discussed above. Thus, in this
case the potential picture does apply for the evaluation of the energy shifts.

On the other hand, if $R\gg a/\alpha$ then $\omega_{char}$ in the energy
 denominator of Eq. (\ref{so}) can be neglected and \be
\delta E_0~\sim~\alpha a^2({\bf E})^2{a\over \alpha}~\sim~{a^3\over R^4}, ~~
~~R
\gg a/\alpha \label{new}
\ee
where we used ${\bf E}^2\sim R^{-4}$ for the characteristic frequencies.
 Eq. (\ref{new}) is
in clear violation of the potential picture. Eq (\ref{new}) could be interpreted by saying that
 the
 electrostatic potential $\sim r^2$ is replaced, when the distance between the
 particles $a$ is much smaller than $\alpha R$, by an effective
potential $\sim r^3$. But this is true only as far as rough estimates are
concerned.
 Rigorously speaking, there is no potential whatsoever corresponding to the
 shifts obtained in this way \cite{voloshin}. Note also that the emergence of
the scale $R\sim a/\alpha$ can be understood in a simple way as an effect of
retardation. Indeed, the time needed to communicate with the distances 
of order of the size of cavity $R$ can be called the retardation time,
$T_{ret}\sim R$. For
the potential picture to be valid this time should be smaller
than the revolution time which is order,
$T_{rev}\sim a/v~\sim a/\alpha$. The potential picture becomes distorted
once $T_{ret}\approx T_{rev}$.

In the QCD case
one considers \cite{voloshin} atom-like systems of heavy quarks $Q$ with a size 
which is, at least formally, much
smaller than the confinement radius,
\be
\alpha_s(M)\cdot M~\gg~ \Lambda_{QCD} \label{ineq}
\ee
where $M$ is the heavy quark mass. In this case the quarks are turning so fast and at such a small
distance that they cannot be resolved by soft gluons. To account for the interaction of
 the $Q \bar {Q}$ pair with these soft gluons one starts again with
the interaction:
\begin{equation}
H_{int}~=~ - \sqrt{\alpha_s}(t_1^a-t_2^a){\bf d}\cdot {\bf E}^a,  \label{dipoleqcd}
\end{equation}
where, the $t_i^a$ refer to the $Q$ and ${\bar Q}$ in the quarkonium, and by ${\bf E}^a$ 
one understands the soft gluonic fields (vaccum fields). While the exact form
of these fields is an as-yet unsolved strong-coupling problem, one
assumes usually \cite{svz} that there is a typical frequency,
$\omega_{char}\sim \Lambda_{QCD} $. The intensity of the vacuum
fields is characterized by a vacuum expectation
value $\langle \alpha_s (G_{\mu\nu}^a)^2 \rangle$ which is treated
phenomenologically.

It may be worthwhile here to discuss the situation represented by Eq. (\ref{ineq})  from the
renormalon viewpoint. In order to do this we consider the shift in the
interaction energy of the quarks in the quarkonium and look for the
infrared sensitive contributions in perturbation theory.  This is seen to arise from the class of 
diagrams shown in Fig.(1a). The external states are the color singlet quarkonium states, the
exchanged gluon is soft (renormalon chain) and the intermediate state in this figure is a continuum
color octet state which is short lived because of its high virtuality. Indeed the energy denominator
for this intermediate state is
$\sim$ $(E-\epsilon)$ $\sim$ $B$, where $E$ is its energy, and
$\epsilon$ that of the external color singlet quarkonium state with binding energy $B$ .
Since $B \gg \Lambda_{QCD}$, (Eq.(\ref{ineq})), we see that the intermediate state is far off shell
and we thus arrive at the reduced diagram shown in Fig (1b). The renormalon chain on the soft gluon
line in this figure is the standard representation of the ( perturbative ) gluon
condensate $<\alpha_s G^2>$ which drives this contribution. A standard analysis ( see e.g.,
\cite{casimir} ) produces a contribution proportional to $\Lambda_{QCD}^4r^3$.  It should be
emphasized that the situation being discussed, i.e., $T_{rev}\ll\Lambda_{QCD}^{-1}$,is precisely the
case when the potential picture is not valid for the non coulombic corrections. 

Both the cases of static quarks \cite{balitsky,simonov} and of atom-like
 systems (quarkonium) \cite{voloshin}
with $E_0\gg \Lambda_{QCD}$ were considered in the literature.
In the latter case one arrives at a natural generalization of Eq. (\ref{new}):
\be
\delta E~\sim n^6\,{{\Lambda_{QCD}^4}\over{(m\alpha_s)^4}}m\label{four} \ee
where we have indicated also the sharp dependence on the principal quantum number $n$.
Note also that Eq. (\ref{four}) has one extra power of $\alpha$ is the
denominator as compared with (\ref{new}) because it is the product
$\alpha_s(G^a_{\mu\nu})^2$ which is renormalization group invariant.

It is worth emphasizing also that in the QCD case there exists an extra problem
 with applying the potential picture to the bound $Q\bar{Q}$ states. Namely,
the energies of intermediate states in Eq. (\ref{so}) are now energies of color
 states since gluons carry color. Formally, these are $Q\bar{Q}$ continuum
states.
Color states are of course widely used in perturbative QCD, and in this respect
the situation does not look exceptional. The standard constraint, however,
is that color states can be excited only for short times and Eq. (\ref{so}),
 when applied in QCD, is therefore consistent as long as the energy denominator
 is much larger than $\Lambda_{QCD}$, i.e., in the situation discussed above.

These considerations suggest that, to be rigorous, the only case which can be
 treated consistently is when
\be
E_n~\sim~M{\alpha_s(M)^2 \over n^2}~\gg ~\Lambda_{QCD} \ee
This implies that it is not just the size of the system that has to be smaller
 than $\Lambda_{QCD}$, but that the binding energy has to be large compared 
to $\Lambda_{QCD}$. In
the real world, where the condition is never satisfied by a wide margin,
only a detailed calculation can tell the extent and degree of the applicability
 of the formalism. In this respect it would seem that reasonably good
results may
be obtained for $\bar{b}b$ with $n=1$ and, to a lesser extent, $n=2$ and
$\bar{c}c$
 with $n=1$ \cite{yndurain}.

Returning to the case of the linear correction to the potential (\ref{novel})
we observe that for large enough quark masses, the retardation effects wipe 
out any kind of a potential
so that the Voloshin-Leutwyler regime sets in. However, for the linear correction 
it happens at larger masses than for the quadratic correction since it is associated
with shorter distances $R_{cr}\sim \alpha_s^{1/2}a^{1/3}\Lambda_{QCD}^{-2/3}$ (see (\ref{ratio}))
where $a$ is the Bohr radius. Therefore, the retardation and revolution times 
get comparable if
\be
a~\le~\alpha^{9/4}\Lambda^{-1}_{QCD}
\ee
where $a$ is understood to be a function of the mass (and of the quantum number
$n$). There is an extra power of $\alpha $ on the right hand side compared to   
the case of the standard correction.

{\bf 9.} To summarize, we have argued that the assumption that weak color 
fields do not penetrate the vacuum, implies a linear correction to the static
quark potential at short distances. For this hypothesis to be realized there
should exist small-size 
non-perturbative fluctuations. If these fluctuations are monopoles then the
linear potential at large distances could possibly be extrapolated to short
distances which does not contradict any (lattice) data at the moment.
The recent indication 
\cite{marchesini} that the running coupling
has a $1/Q^2$ correction does imply a linear correction to the potential
as well. If confirmed, therefore, it would indicate the
existence of small-size non-perturbative fluctuations. We have also considered 
phenomenological implications of the linear correction to the potential
for quarkonia but found that at this time no conclusion can be drawn on 
the existence or absence of such a correction.  

 \vspace{1.0cm}
{\bf Acknowledgements}. Part of this work was done in April 1995 when
we were visiting the University Autonoma de Madrid. We are grateful to the
 Departamento de F\'\i sica Te\'orica there for hospitality. We would like to thank F. J. Yndur\'ain 
for an initial collaboration.  We have benefitted from interesting discussions with G. Marchesini, A.
Vainshtein, P. van Baal.

\newpage

\[ \epsfxsize=7.0cm \epsfbox{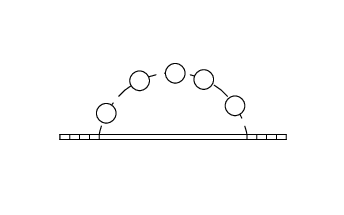} \]
\vspace{-.5cm}
\[ \parbox{10cm}{ {\bf Fig. 1a} Class of diagrams contributing to the shift in 
interaction energy of heavy quarks in quarkonium} \]

\[ \] \[ \]

\[ \epsfxsize=7.0cm \epsfbox{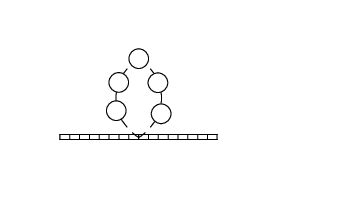} \]
\vspace{-.5cm}
\[ \parbox{10cm}{ {\bf Fig. 1b}  Reduced diagram for far off-shell 
  intermediate states. } \]

\end{document}